\documentclass[rnote,longauth]{aa} 
\usepackage{natbib}
\bibpunct{(}{)}{;}{a}{}{,} 
\usepackage{graphicx}
\usepackage{txfonts}

\def\4U{4U\,0142+61}
\def\1e{1E\,2259+586}
\begin{document}
   \title{Observations of the magnetars \4U and \1e with the MAGIC telescopes}

%
\author{
 J.~Aleksi\'c\inst{1} \and
 L.~A.~Antonelli\inst{2} \and
 P.~Antoranz\inst{3} \and
 M.~Asensio\inst{4} \and
 U.~Barres de Almeida\inst{5} \and
 J.~A.~Barrio\inst{4} \and
 J.~Becerra Gonz\'alez\inst{6,}\inst{28} \and
 W.~Bednarek\inst{7} \and
 K.~Berger\inst{6,}\inst{8} \and
 E.~Bernardini\inst{9}
% ...
}
\institute { IFAE, Edifici Cn., Campus UAB, E-08193 Bellaterra, Spain
 \and INAF National Institute for Astrophysics, I-00136 Rome, Italy
 \and Universit\`a  di Siena, and INFN Pisa, I-53100 Siena, Italy
 \and Universidad Complutense, E-28040 Madrid, Spain
 \and Max-Planck-Institut f\"ur Physik, D-80805 M\"unchen, Germany
 \and Inst. de Astrof\'{\i}sica de Canarias, E-38200 La Laguna, Tenerife, Spain
 \and University of \L\'od\'z, PL-90236 Lodz, Poland
 \and Depto. de Astrof\'{\i}sica, Universidad de La Laguna, E-38206 La Laguna, Spain
 \and Deutsches Elektronen-Synchrotron (DESY), D-15738 Zeuthen, Germany
 \and ETH Zurich, CH-8093 Zurich, Switzerland
 \and Universit\"at W\"urzburg, D-97074 W\"urzburg, Germany
 \and Centro de Investigaciones Energ\'eticas, Medioambientales y Tecnol\'ogicas, E-28040 Madrid, Spain
 \and Universit\`a di Udine, and INFN Trieste, I-33100 Udine, Italy
 \and Institut de Ci\`encies de l'Espai (IEEC-CSIC), E-08193 Bellaterra, Spain
 \and Technische Universit\"at Dortmund, D-44221 Dortmund, Germany
 \and Croatian MAGIC Consortium, Rudjer Boskovic Institute, University of Rijeka and University of Split, HR-10000 Zagreb, Croatia
 \and Universitat Aut\`onoma de Barcelona, E-08193 Bellaterra, Spain
 \and Tuorla Observatory, University of Turku, FI-21500 Piikki\"o, Finland
 \and Japanese MAGIC Consortium, Division of Physics and Astronomy, Kyoto University, Japan
 \and Inst. for Nucl. Research and Nucl. Energy, BG-1784 Sofia, Bulgaria
 \and Universitat de Barcelona (ICC/IEEC), E-08028 Barcelona, Spain
 \and Universit\`a di Padova and INFN, I-35131 Padova, Italy
 \and INAF/Osservatorio Astronomico and INFN, I-34143 Trieste, Italy
 \and Universit\`a  dell'Insubria, Como, I-22100 Como, Italy
 \and Inst. de Astrof\'{\i}sica de Andaluc\'{\i}a (CSIC), E-18080 Granada, Spain
 \and Universit\`a  di Pisa, and INFN Pisa, I-56126 Pisa, Italy
 \and ICREA, E-08010 Barcelona, Spain
 \and now at: Institut f\"ur Experimentalphysik, Universit\"at Hamburg, Germany
% ...
}

\offprints{corresponding author D.~Hadasch (hadasch@ieec.uab.es)}


 
  \abstract
   {Magnetars are an extreme, highly magnetized class of isolated neutron stars whose large X-ray luminosity is believed to be driven by their high magnetic field.}
   {Study for the first time the possible very high energy $\gamma$-ray emission above 100\,GeV from magnetars, observing the sources \4U and \1e.}
   {We observed the two sources with atmospheric Cherenkov telescopes in the very high energy range (E$>100$\,GeV). 
\4U was observed with the MAGIC I telescope in 2008 for $\sim$25\,h and \1e was observed with the MAGIC stereoscopic system in 2010 for $\sim$14\,h. The data were analyzed with the standard MAGIC analysis software.}
   {Neither magnetar was detected. Upper limits to the differential and integral flux above 200\,GeV were computed using the Rolke algorithm. We obtain integral upper limits to the flux of 1.52$\times10^{-12}$cm$^{-2}$ s$^{-1}$ and 2.7$\times10^{-12}$cm$^{-2}$ s$^{-1}$ with a confidence level of 95\% for \4U and \1e, respectively. The resulting differential upper limits are presented together with X-ray data and upper limits in the GeV energy range.}
{}

   \maketitle
%

\section{Introduction}

Magnetars are a peculiar class of neutron stars.
Most of the about 20 known magnetars are characterized by strong dipolar magnetic fields ($\sim10^{14}-10^{15}$\,Gauss) that are $\sim10-1000$ times higher than the average value in radio pulsars, near or even above the quantum electrodynamic field strength, $B_{\mathrm{QED}} = m^{2}_{e}c^{3}/e\hbar \sim 4.4\times 10^{13}$\,G \citep{harding2006}, although with two exceptions \citep{rea2010,rea2012}.
They have bright X-ray luminosities
($L_{\mathrm{x}}\sim10^{32}-10^{36}$\,erg\,s$^{-1}$) from 0.1--300\,keV, longer rotation periods than most ordinary radio pulsars ($\sim2$--$12$\,s),
and very high period derivatives ($\sim 10^{-13}$--$10^{-11}$\,s/s).
For more details see recent reviews on magnetars by \citet{mereghettiReview} and \citet{reaReview}.

The most successful model for explaining the X-ray emission from these objects invokes the decay and instability of their magnetic fields \citep{duncan1992, thompson1993, thompson1995}.
The dichotomy between magnetars and ordinary pulsars may indicate different progenitors \citep{duncan1996}.
This scenario for birthing a magnetar postulates a very rapidly
spinning proto-neutron star at birth, which would then have a high rotational
energy. 
This excessive energy was searched in the supernova remnants (SNRs)
surrounding magnetars.
However, those SNRs show no excess in X-rays relative to those around normal pulsars \citep{vink2006}.

With this paper we aimed at testing
whether the additional energy at birth could have gone in TeV
emission.

To date, no magnetar has been detected at energies above 1\,MeV.
Although recently, 
the H.E.S.S. collaboration presented their discovery of
extended TeV $\gamma$-ray emission towards the magnetar SGR 1806-20\footnote{Originally, magnetars were divided into two categories: Soft Gamma Repeaters (SGRs) and Anomalous X-ray Pulsars (AXPs) \citep{AXPSGRReview}.}, it is doubtful that the emission is driven by
the magnetar itself \citep{sgr1806hess}.
The \textit{Fermi}-LAT Collaboration presented upper limits for 13 magnetars after 17 months of sky survey observations between 0.1 and 100\,GeV \citep{fermimagnetar}.
\cite{mus2010} studied specially the Fermi data of \4U. 
Neither steady nor pulsed emission was found.
In this work we present a search for the emission
at very high energies (200\,GeV--50\,TeV) from the two
magnetars \4U and \1e with the MAGIC
telescopes. These sources
have been also observed by the VERITAS Collaboration and corresponding
upper limits above an energy of 400\,GeV have
been presented in \cite{veritas2009}.
The present MAGIC
observations of these two magnetars extend the spectrum to lower energies, 200GeV.

\section{The observed magnetars}

The source \4U is located at $\alpha_{2000}, \delta_{2000}= \rm 01^h46^m22\fs407, +61\degr45' 03\farcs19$ at a distance of $\sim3.6$\,kpc.
With an X-ray luminosity of $L_{\mathrm{X}}\sim1\times 10^{35}\mathrm{erg}$\,$\mathrm{s}^{-1}$
it is one of the most X-ray luminous magnetars known \citep{mcgill}.
This makes it a good target to search for persistent very high energy
emission.
Long term spin period variations ($P\sim8.7$\,s) were discovered during observations with \textit{EXOSAT} \citep{israel1994}, leading to the measurement of
the period derivative $\dot P$
$\sim2\times10^{-12}\mathrm{s}\mathrm{s}^{-1}$, and consequently of the very
strong magnetic field $B\sim1.3\times10^{14}\mathrm{G}$ \citep{mcgill}. The
bright 1-10\,keV emission coming from \4U has been observed by many X-ray satellites \citep{white1987, israel1999, patel2003, rea2007a, rea2007b} revealing
an X-ray spectrum typical of an Anomalous X-ray Pulsar (AXP), best described by an absorbed blackbody plus
a power law ($N_{\mathrm{H}} \sim10^{22}\mathrm{cm}^{-2}, kT\sim0.4
\mathrm{keV}$ and $\Gamma\sim3.62$). A very strong hard
X-ray emission has been reported by \textit{INTEGRAL} up to 250\,keV, with a spectrum
well modeled with a steep power-law with a photon index of $\sim$1 \citep{kuiper2006}.
At the time of data taking with the MAGIC telescope, there were only \textit{COMPTEL} upper limits in the MeV range suggesting a spectral break in the hard
X-ray emission of this object.
The upper limits, however,
do not put strong constraints on the HE or VHE gamma-ray emission of
the object, especially given the high 
systematic uncertainty of the background 
subtraction in the data COMPTEL analysis \citep{comptelbkg}.
Recently, the upper limits derived by the \textit{Fermi}-LAT Collaboration \citep{fermimagnetar} and by \cite{mus2010} point to a cutoff in the MeV band.

The AXP \1e is located at $\alpha_{2000}, \delta_{2000}= \rm 23^h01^m08\fs296, +58\degr52'44\farcs45$ embedded in the SNR CTB109. The source has a magnetic field of $B\sim0.59\times10^{14}\mathrm{G}$ and a distance of $\sim$4\,kpc, making it a good candidate
for MAGIC observations \citep{mcgill}.
\textit{RXTE} measured the period ($P\sim7$\,s) and the period derivative
($\dot P\sim0.5\times10^{-12}\mathrm{s}\mathrm{s}^{-1}$) \citep{gavriil2002}.
The X-ray spectrum is variable depending on the source emission state \citep{kaspi2003,woods2004}. 
After undergoing an outburst in 2002, the source returned into its possible quiescence state and the corresponding spectrum is best fitted by a blackbody plus a power law ($N_{\mathrm{H}} \sim10^{22}\mathrm{cm}^{-2}, kT\sim0.4\mathrm{keV}$ and $\Gamma\sim3.75$) \citep{zhu2008}.

\section{The MAGIC telescopes, analysis and data}

The MAGIC Collaboration operates two 17\,m diameter imaging atmospheric Cherenkov
telescopes on the Canary Island of La Palma. The data sets presented here were
taken in 2008, i.e. 
before the second MAGIC telescope was operational (mono data), and in
2010 when both telescopes were already taking stereoscopic data.
Details about the performance of MAGIC in mono and stereo mode can be found in 
\cite{crabmagicI} and \cite{aleksicanalysis}.
All data presented in this work were taken in the so-called wobble mode
and were analyzed using the MARS analysis framework \citep{moralejo2009,aleksicanalysis}.
The analyses presented here have an analysis threshold of 200\,GeV.
The upper limits were calculated using the Rolke algorithm \citep{rolke} with a confidence level (C.L.) of 95\% assuming a Gaussian
background and 30\% of systematic uncertainty in the flux level.
Since \1e is embedded in a SNR and may contain more than one emission region (see below)
relevant parameters for the observations are the MAGIC field of view of 3.5$^\circ$ and the
angular resolution 
of $\sim$0.07$^\circ$ above 300\,GeV \citep{aleksicanalysis}.

\begin{table*}[tbh]
\begin{center}
\caption{\label{table:mag_table}Magnetar parameters taken from \cite{mcgill}, along with the MAGIC results presented here. Crab Units (C.U.) are defined as a fraction of the Crab Nebula flux as measured by MAGIC \citep{aleksicanalysis}.}
\begin{tabular}{ l c c c c c c c}
\hline \hline
Source          & Distance    & $B_{\mathrm{surf}}$         & $L_{\mathrm{X}}$                       &  log($L_{\mathrm{rot}}$)  &  Eff. obs. time  &  Significance &   Upper limit (95\% C.L., E$>$200\,GeV) \\
                &   [kpc]     &   [10$^{14}$G] &  [10$^{35}$erg s$^{-1}$]      &  [erg s$^{-1}$]  & [hrs]          &     $\sigma$  &    [cm$^{-2}$ s$^{-1}$]    \\ \hline
\4U\            & 3.6$\pm$0.4 & 1.3                & 1.1                           & 32.10            &   16.58        &     $-$2.1      &        1.52$\times10^{-12}$ (0.70\% C.U.)             \\
\1e\            & 4.0$\pm$0.8 & 0.59               & 0.34                          & 31.70            &   8.22         &     $-$0.5      &        2.70$\times10^{-12}$ (1.24\% C.U.)           \\
\hline
\end{tabular}
\end{center}
\end{table*}

\4U was observed for 25.41\,hours. After quality cuts
 16.58\,hours of effective observation time remain.
These mono data were taken between August and December 2008 covering a zenith angle range between 33$^\circ$ and 40.6$^\circ$.

Data for \1e were taken in stereo mode wobbling around the sky position 0.12$^\circ$ away from the magnetar to have the shell of
the supernova remnant and the magnetar in the same field of view.
Given the angular resolution of the MAGIC telescopes,
these two possible TeV sources would be spatially separable
with MAGIC.
The region was observed between
August and November 2010 for 14.33\,hours within a zenith angle range of 29$^\circ$--43$^\circ$.
After quality cuts this amounted to 8.22\,hours of effective observation time.

\section{Results}

\begin{figure}[]
  \begin{center}
    \includegraphics*[width=0.45\textwidth,angle=0,clip]{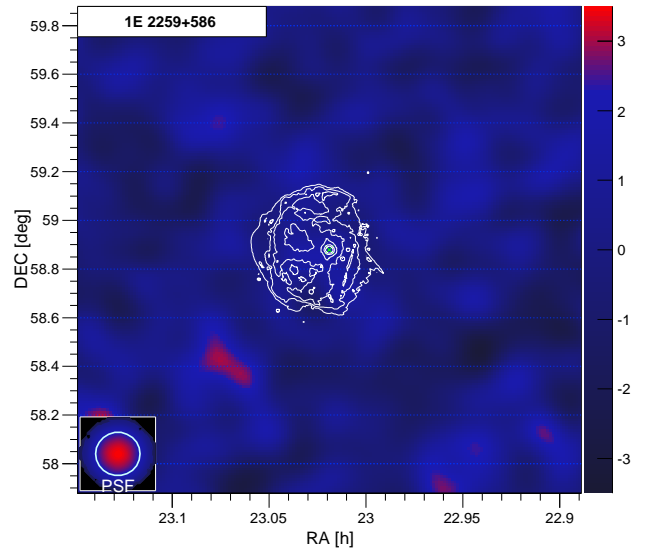}
    \caption{\label{fig:TSmap}TS map of \1e. The green cross represents the magnetar position. The white contours show the X-ray emission of the surrounding SNR CTB 109 detected with the \textit{XMM-Newton} satellite. The color scale represents the TS value.}
  \end{center}
\end{figure}

Neither source was detected by MAGIC. We computed the
integral flux upper limits above 200\,GeV with 95\% C.L. assuming a differential energy spectral shape of a power law
with an index of 2.6, similar to that of the Crab Nebula spectrum.
The results are given in Table~\ref{table:mag_table}.
A 25\% change in the photon index yields a variation of about 7\%.
\enlargethispage{\baselineskip}
In Fig.~\ref{fig:TSmap} we show the corresponding test statistic (TS)\footnote{
Our test statistic is \cite{lima} eq. 17, applied on a smoothed and
modeled background estimation. Its null hypothesis distribution mostly
resembles a Gaussian function, but in general can have a somewhat
different shape or width.} map for \1e.
No excess was found at either the magnetar position nor at any location within the surrounding SNR.
The TS map for \4U is not shown here, but shows the same flat behaviour.
The upper limit for the extended SNR will be discussed elsewhere. The white contours represent the X-ray emission of the surrounding SNR detected with the \textit{XMM-Newton} satellite (0.1--15\,keV).
We also searched for pulsations for both magnetars. For the pulsed
analysis of 1E 2259+586 we used a timing solution valid at the epoch of the MAGIC
observations, as derived by \citet{icdem2012}.
We did not detect any significant pulsation at VHE energies.
In the case of \4U, we searched for pulsation using the ephemeris of
\citet{mus2010}. We did not find any pulsation at VHE energies for this source
either.

\begin{figure}[]
  \begin{center}
    \includegraphics*[width=0.49\textwidth,angle=0,clip]{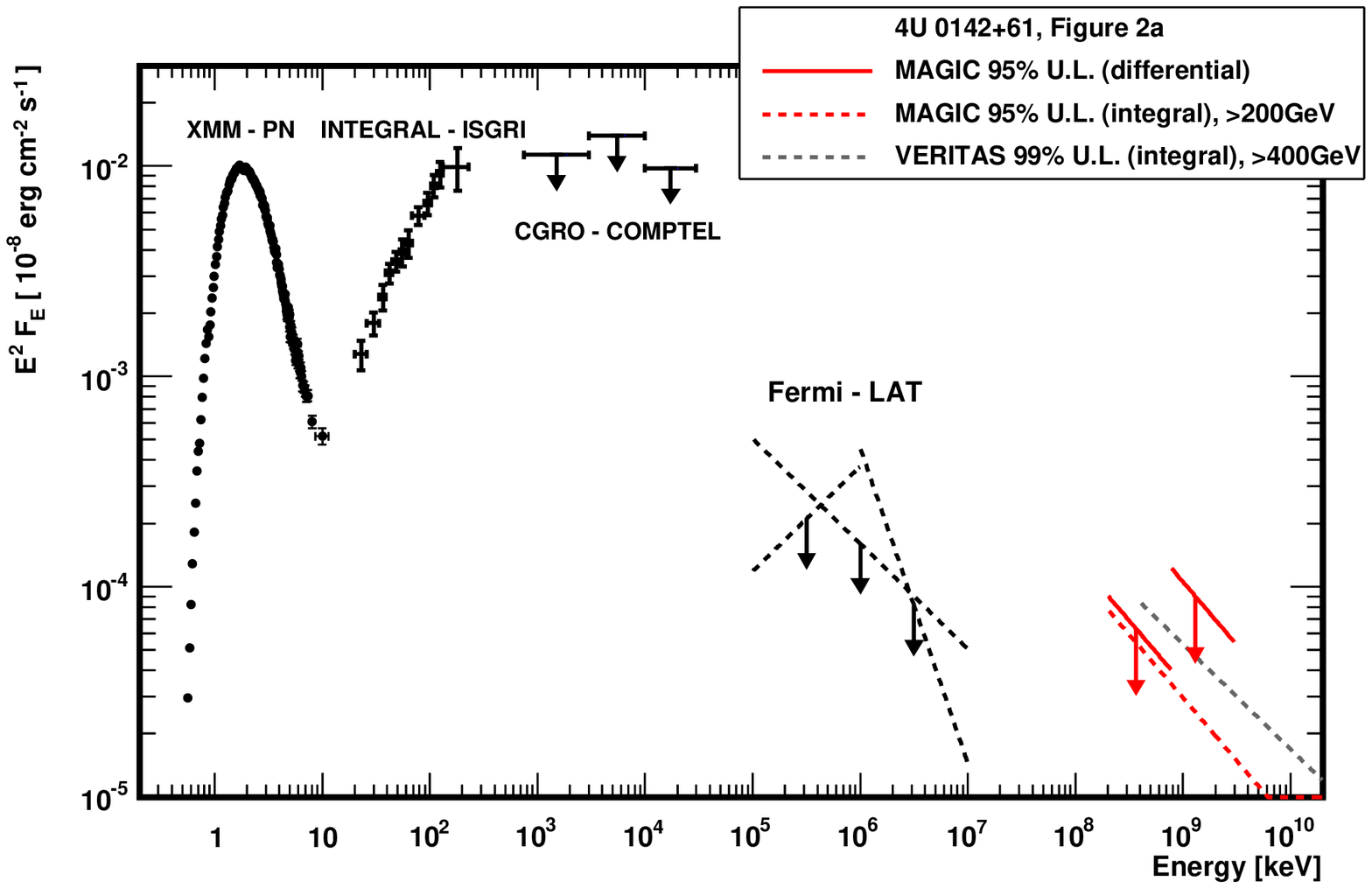}
    \includegraphics*[width=0.49\textwidth,angle=0,clip]{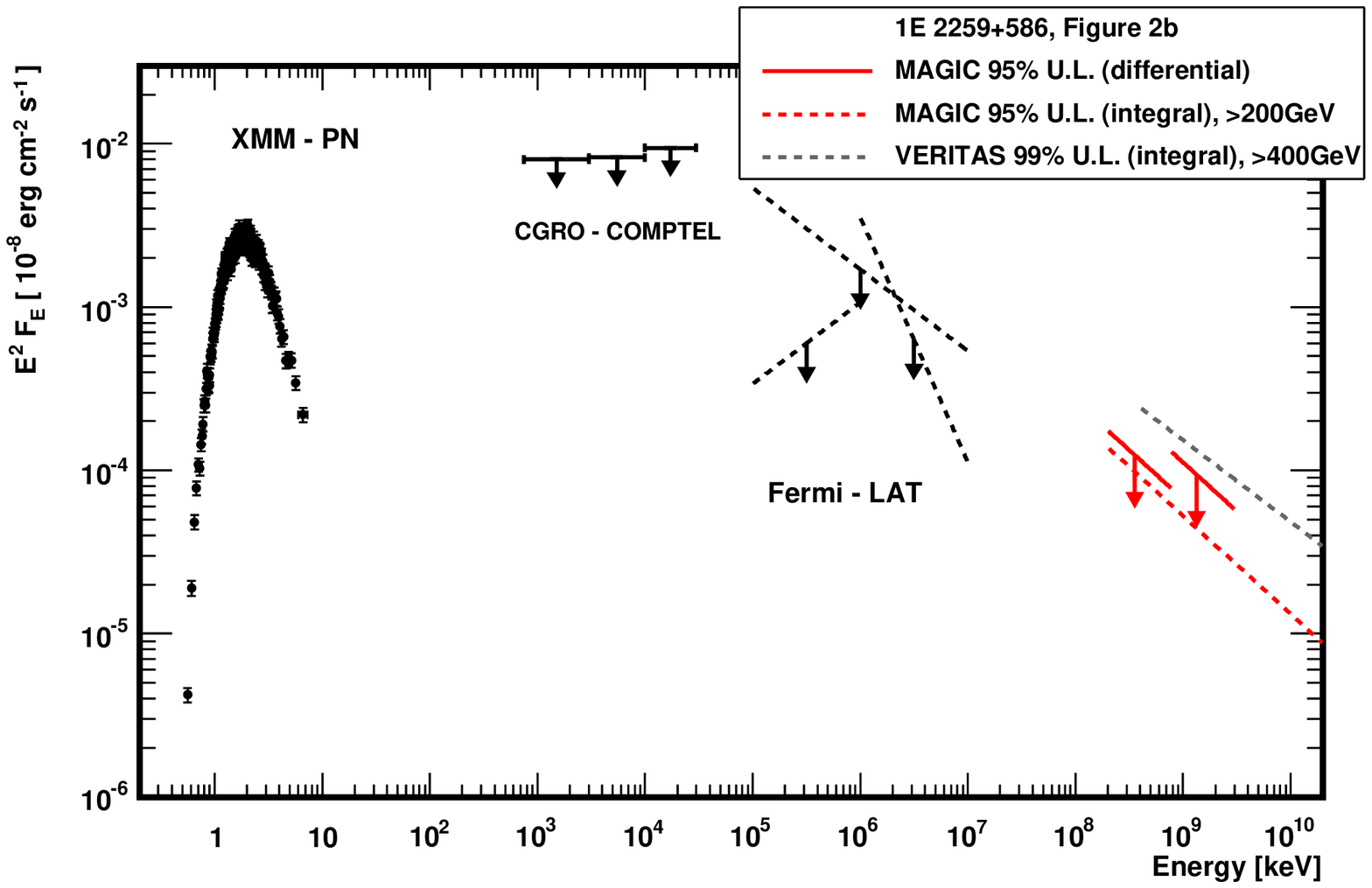}
    \caption{\label{fig:SED_4U}Spectral energy distributions of \4U\ (2a) and \1e (2b) from X-rays to TeV energies. In black the points and upper limits in the keV up to the GeV energy range are shown. The upper limits derived by the VERITAS Collaboration are shown in gray and the upper limits from this work are shown in red. See text for further details on the data and upper limits presented here.}
  \end{center}
\end{figure}

Since neither source experienced an outburst in X-rays during our observing intervals, we can compare our upper limits with data
taken with different instruments during different quiescent epochs.
In Fig.~\ref{fig:SED_4U}a (\ref{fig:SED_4U}b)  we present the 0.1\,keV--3\,TeV multi-band spectral energy distribution (SED) of \4U (\1e), respectively.
For both sources the corresponding differential and integral upper limits derived in this work are shown
(red lines in Fig.~\ref{fig:SED_4U}).
In the case of \4U, the 0.1--200\,keV data are from \textit{XMM}-Newton-PN and \textit{INTEGRAL}-ISGRI \citep{rea2007a,hartog4U2008,gonzalez4U2010} plotted together with the 2$\sigma$ COMPTEL upper limits \citep{hartog2006,kuiper2006}. For \1e we show data points from \textit{XMM}-Newton-PN \citep{woods2004} together with COMPTEL upper limits \citep{kuiper2006}.
The upper limits provided by the \textit{Fermi}-LAT Collaboration were calculated for three different energy ranges \citep{fermimagnetar}. For the overall energy bin from 0.1--10\,GeV a photon index of 2.5 was assumed.
A cutoff is mimicked by splitting this energy bin into two parts with photon indices of 1.5 and 3.5, respectively. The assumed slopes are indicated in Fig.~\ref{fig:SED_4U}.
The results derived by the VERITAS Collaboration on the two sources are also shown
for comparison (grey dashed lines). They correspond to 99\% C.L. integral flux upper limits of 
$8.68\times10^{-13}\mathrm{cm}^{-2}\mathrm{s}^{-1}$ for \4U and 
$2.49\times10^{-12}\mathrm{cm}^{-2}\mathrm{s}^{-1}$ for \1e by assuming a power-law with a photon index of 2.5 above 400\,GeV \citep{veritas2009}.
The upper limits for both sources are compatible with a
break in the power law at $\sim$1\,MeV. However, the SED lacks any measurements above hard X-rays, what
gives complete freedom under the corresponding instrumental
sensitivity.

\cite{cheng2001} presented a model for the very high energy radiation from magnetars.
They predicted emission of $\gamma$-rays in the GeV band coming from
the outer gap for the two sources we studied.
This model has been recently revised by \cite{tong2011}, who updated the observational parameters to
calculate the $\gamma$-ray radiation properties of all AXPs and SGRs using the models by
\cite{zhang1997} and \cite{cheng2001}.
The scenario by \cite{tong2011} predicts
that \4U should have been detected by
\textit{Fermi}-LAT, although they explain the lack of a detection by Fermi-LAT \citep{fermimagnetar,mus2010} by invoking accretion.
For \1e the model does not predict GeV emission.
We note that although none of the current models predict TeV range emission for either magnetar, the existence of diffuse emission around \1e could lead to the appearance of an extra
component in the SED besides any magnetospheric emission. 

Using the MAGIC telescopes we studied for the first time the possibility of magnetars to be a new TeV source class on the examples of \4U and \1e.
This exploratory work led to a non-detection of the VHE gamma-ray emission from
either of them. This result indicates that magnetars are probably not VHE
emitters during their quiescent state, as expected from the various theoretical
models. However, the possibility of magnetars being VHE emitters during
flaring episodes cannot be ruled out because of the lack of VHE observations
during these high-activity periods. Consequently, our future searches for VHE
emission of magnetars will be performed during outbursts 
\footnote{In order to provide fast reactions to such events in the future,
MAGIC has installed an alert system, which receives alerts provided by several satellites and points the telescopes
to the flaring source automatically, as it is also done for observations of Gamma Ray Bursts.}.

\begin{acknowledgements}
We would like to thank the Instituto de Astrof\'{\i}sica de
Canarias for the excellent working conditions at the
Observatorio del Roque de los Muchachos in La Palma.
The support of the German BMBF and MPG, the Italian INFN, 
the Swiss National Fund SNF, and the Spanish MICINN is 
gratefully acknowledged. This work was also supported by the CPAN CSD2007-00042 and MultiDark
CSD2009-00064 projects of the Spanish Consolider-Ingenio 2010
programme, by grant DO02-353 of the Bulgarian NSF, by grant 127740 of 
the Academy of Finland,
by the DFG Cluster of Excellence ``Origin and Structure of the 
Universe'', by the DFG Collaborative Research Centers SFB823/C4 and SFB876/C3,
and by the Polish MNiSzW grant 745/N-HESS-MAGIC/2010/0.
\end{acknowledgements}

\bibliographystyle{aa}
\bibliography{/Users/hadasch/Paper/MyPaper/MAGICmagnetars/ver_arxiv_corrected/bibtex/bibtex_magnetars}

\end{document}